\documentclass[twocolumn,aps,prc,showpacs,superscriptaddress,preprintnumbers,floatfix,nofootinbib]{revtex4}
\usepackage{epsfig,graphics}
\usepackage{graphicx}
\usepackage{dcolumn}
\usepackage{bm}
\usepackage{amsmath}
\usepackage[usenames]{color}
\usepackage{ulem} 

\begin{document}

\title{ Reaction plane angle dependence of dihadron azimuthal correlations from a multiphase transport model calculation }

\author{ W. Li}
\affiliation{Shanghai Institute of Applied Physics, Chinese
Academy of Sciences, P.O. Box 800-204, Shanghai 201800, China}
\affiliation{Graduate School of the Chinese Academy of Sciences,
Beijing 100080, China}
\author{ S. Zhang}
\affiliation{Shanghai Institute of Applied Physics, Chinese
Academy of Sciences, P.O. Box 800-204, Shanghai 201800, China}
\affiliation{Graduate School of the Chinese Academy of Sciences,
Beijing 100080, China}
\author{ Y. G. Ma}
\thanks{Corresponding author: Email: ygma@sinap.ac.cn}
\affiliation{Shanghai Institute of Applied Physics, Chinese
Academy of Sciences, P.O. Box 800-204, Shanghai 201800, China}
\author{ X. Z. Cai}
\affiliation{Shanghai Institute of Applied Physics, Chinese
Academy of Sciences, P.O. Box 800-204, Shanghai 201800, China}
\author{ J. H. Chen}
\affiliation{Shanghai Institute of Applied Physics, Chinese
Academy of Sciences, P.O. Box 800-204, Shanghai 201800, China}
\author{H. Z. Huang}
\affiliation{ Dept of Physics and Astronomy, University of
California at Los Angeles, CA 90095, USA}
\author{ G. L. Ma}
\affiliation{Shanghai Institute of Applied Physics, Chinese
Academy of Sciences, P.O. Box 800-204, Shanghai 201800, China}
\author{ C. Zhong}
\affiliation{Shanghai Institute of Applied Physics, Chinese
Academy of Sciences, P.O. Box 800-204, Shanghai 201800, China}

\date{ \today}
\begin{abstract}

Dihadron azimuthal angle correlations relative to the reaction
plane have been investigated in Au + Au collisions at
$\sqrt{s_{NN}}$ = 200 GeV using a multi-phase transport model
(AMPT). Such reaction plane azimuthal angle dependent correlations
can shed light on path-length effect of energy loss of high
transverse momentum particles propagating through the hot dense
medium. The correlations vary with the trigger particle azimuthal
angle with respect to the reaction plane direction,
$\phi_{s}=\phi_{T}-\Psi_{EP}$, which is consistent with the
experimental observation by the STAR collaboration. The dihadron
azimuthal angle correlation functions on the away side of the
trigger particle present a distinct evolution from a single peak
to a broad, possibly double peak, structure when the trigger
particle direction goes from in-plane to out-of-plane of the
reaction plane. The away-side angular correlation functions are
asymmetric with respect to the back-to-back direction in some
regions of $\phi_{s}$, which could  provide insight on testing
$v_{1}$ method to reconstruct the reaction plane. In addition,
both the root-mean-square width ($W_{rms}$) of the away-side
correlation distribution and the splitting parameter $D$ between
the away-side double peaks increase slightly with $\phi_{s}$, and
the average transverse momentum of the away-side associated
hadrons shows a strong $\phi_{s}$ dependence. Our results indicate
that strong parton cascade and resultant energy loss could play an
important role for the appearance of a double-peak structure in
the dihadron azimuthal angular correlation function on the away
side of the trigger particle.

\end{abstract}

\pacs{12.38.Mh, 11.10.Wx, 25.75.Dw}

\maketitle
\section{Introduction}

Lattice QCD calculations predicted that a novel state of
matter~\cite{QCD}, distinct from ordinary hadronic matter, can be
created in ultra-high energy heavy-ion collisions. The BNL
Relativistic Heavy Ion Collider (RHIC) was constructed to
investigate the properties of the new state of
matter~\cite{White-papers}. Studying the properties of this dense
nuclear matter is a great challenge both experimentally and
theoretically. Recently features of partonic collectivity and
strongly interacting have been found for the matter of extreme
temperature and energy density created in nucleus-nucleus
collisions at RHIC~\cite{ellipticflow,STARphi,jpsi,jet-ex}. The
energy loss of hard partons traversing the dense medium created in
the collisions is a crucial question. Usually, there are two main
mechanisms  causing a jet to lose energy: elastic collisions with
deconfined partons and induced gluon
radiation~\cite{jet-quenching1,jet-quenching2,energy_loss,Zapp3,elastic-collision1,inelastic1,inelastic2},
and the gluon radiation is considered as the main process
responsible for energy loss. Both mechanisms suggest that the
energy loss would depend strongly on the traversed path length of
the propagating
jets~\cite{elastic-collision2,elastic-collision3,soft-soft-th2}.
Two particle azimuthal angular correlation has been demonstrated
as a good probe to investigate the interactions between jets and
the hot dense medium~\cite{hard-hard-ex,soft-soft-th1}. It has
been reported previously, at RHIC, that for a high $p_{T}$ trigger
particle, the correlated yields at $p_{T}>$2 $ GeV/c$ on the away
side are strongly suppressed~\cite{hard-hard-ex}, while for lower
$p_{T}$ the yields are enhanced. The correlated hadrons on the
away-side appear to be partially equilibrated with the bulk medium
and the azimuthal angular correlation distributions are much
broader, possible with double
peaks~\cite{soft-soft-ex1,sideward-peak1,sideward-peak3}. Recently
many theoretical models have been proposed to describe the
phenomenon$\colon$ Mach-cone shock waves generated by large energy
deposition in the medium~\cite{Casalderrey}, conical emission due
to Cherenkov radiation~\cite{Koch}, the excitation of collective
plasmon waves~\cite{Ruppert}, large angle gluon
radiation~\cite{large-angle,opaque-media-radiation}, strong parton
cascade~\cite{di-hadron,three-hadron,time-evolution}, jets
deflected by radial flow~\cite{Ruppert,deflected-jets} and
path-length dependent energy loss~\cite{path-length-dependent}. It
has also been observed previously that the double peak structure
in the dihadron angular correlation strongly depends on
$\phi_{s}$, the trigger particle azimuthal angle with respect to
the direction of the event plane~\cite{Wang,aoqi}. The angular
correlation distributions on the away side are different between
in-plane and out-of-plane orientations which indicate a strong
path-length effect on parton-medium interactions~\cite{aoqi}.
However, no Monte Carlo simulation has been performed to
demonstrate such kind of reaction plane angle dependence so far.
In this work, we will use a multi-phase transport model
(AMPT)~\cite{AMPT} to reveal the dependence of the reaction plane
orientation on the dihadron correlation structure which can be
understood by the path-length effect of large elastic partonic
energy loss.

AMPT model is a hybrid model which consists of four main
components: the initial conditions, partonic interactions, the
conversion from partonic matter to hadronic matter and hadronic
rescattering. AMPT model includes two versions: the default version and
the string melting version where strings are melted and hadronizations using
parton recombination mechanism.
Previous studies have shown that the partonic effect could not be
neglected and the string melting AMPT model which contains stronger
parton cascade is much more appropriate than the default AMPT
model when the energy density is much higher than the critical
density for the predicted phase
transition~\cite{AMPT,SAMPT,Jinhui}. The parton cascade model ZPC
in the AMPT model only includes elastic 2$\rightarrow$2 partonic
interactions~\cite{AMPT}, higher order inelastic process, which
might become dominant at high densities during the early stage of
relativistic heavy ion collisions \cite{2->3_gluon_radiation2},
have not been included. To fully model the parton energy loss, in addition to
the elastic energy loss the
radiation energy loss (many-body inelastic interactions in
general)~\cite{energy_loss,Zapp3} should also be implemented in this
hybrid transport model. However, it is not the intention of our current work to
embark on such a comprehensive task of implementing radiation energy loss in AMPT.
In order to match the parton energy loss in experimental data a large elastic scattering cross
section ($\sigma$) of 10 mb has to be used in AMPT model.
The large elastic cross section could contribute to the emergence of a double peak structure
on the away side angular correlation distribution.

For mid-central collisions, the initial overlap region of the two
nuclei looks like an almond, Figure 1 shows the initial parton
coordinate distribution for 40\% collision centrality from AMPT model
simulations.  The dimensions of dense matter distribution for in-plane and
out-of-plane are quite different in mid-central collisions. We have
calculated the eccentricity of the initial state, right after the
collision in AMPT model, to be around 0.15.
Particles along out-of-plane will on average propagate a longer
path length than those along the in-plane direction. Triggered high momentum particles are biased preferentially
from the surface area and outward from different azimuthal angles. The corresponding away side jet and the
associated particles will propagate
through different dense medium length. This can result in
different energy loss due to dependence on interaction strength
between the traversing parton and the
dense medium, which may lead to
different correlation structures on the back-to-back direction.

\section{Analysis method}

In order to investigate the path-length effect, we divide the
whole initial angular phase space region into 16 equal slices
based on $\phi_{s}=\phi_{T}-\Psi_{EP}$ in Fig 2,  where $\phi_{T}$
and $\Psi_{EP}$ are the azimuthal angle of triggered particle and
event plane, respectively. We calculate the dihadron azimuthal
angle correlations between a high $p_{T}$ hadron (trigger hadron)
and low $p_{T}$ one (associated hadron) separately in each slice.
The analysis method is similar to that used by the
experiments~\cite{soft-soft-ex,sideward-peak2,flow_construct_1,flow_construct_2}.
In the same event, pairs of an associated particle with a
triggered particle are accumulated to obtain $\Delta\phi$ =
$\phi_{assoc}$ - $\phi_{trig}$ raw distributions, where
$\phi_{assoc}$ and $\phi_{trig}$ are the azimuthal angle of
associated particle and triggered particle, respectively. In order
to reproduce the background which is mainly from the anisotropic
flow [13, 15], a mixing-event method is applied. We mixed the
events which have very close centrality into a new event, and
extracted $\Delta\phi$ distribution to be used as a background
distribution. A zero yield at minimum (ZYAM) assumption was
adopted to subtract the background ~\cite{sideward-peak2}. We can
directly use the event plane angle $\Psi_{EP}$ in our AMPT model
simulation which sets $\Psi_{EP}=0$.

\section{Results and Discussions}

Figure 3 shows the dihadron correlation functions for 16 slices as
determined by the trigger particle emission angle with respect to
the reaction plane. The data are Au+Au collisions  at
$\sqrt{s_{NN}}$ = 200 GeV from AMPT model simulation with a
20-60$\%$ collision centrality. We set $p_{T}$ range for trigger
particles to be $2.5 <p_{T}^{trig}< 6$ GeV/$c$ and for associated
particles to be $0.15 <p_{T}^{asso}< 3$ GeV/$c$ in our analysis.
Both triggered and associated particles are further selected with
a pseudo-rapidity cut of $|\eta|<1$. We can observe distinct
evolutions from in-plane to out-of-plane on both near and away
side. The near side correlations maintain a single Gaussian
structure while the correlation yields decrease from in-plane to
out-of-plane direction. On away-side, however, the correlation
structure evolves from a single peak to double peaks. Furthermore,
we note that the double peaks are not symmetric for data from some
trigger particle emission directions. For further discussion on
the asymmetric structure, we focus on slice 2 and slice 15 as
examples. We find that the asymmetry of their correlation
structures on the away side represents different orientation. For
slice 2 as shown in Fig. 2, the height of peak centered at around
2.8 rad is higher than that centered at around 3.8 rad, which is
due to that on the away side direction particles travel through
longer medium path length at around 3.8 rad (i.e. the left side
medium of the back-side of slice 2, namely slice 10) than that at
2.8 rad (i.e. the right side medium of the slice 10). For slice
15, it shows the opposite asymmetry. To guide the eyes, the two
Gaussian fits are plotted for the Slice 2 and 15. Consequently,
hadrons travel through different medium length and induce
different asymmetry of the away-side dihadron structure. In other
words, the correlation structure depends on the path-length that
the associated particles travel.

Our simulation result seems to indicate that violent parton
cascade processes can make significant contributions to the
splitting structure in the away-side dihadron correlations. Recent
preliminary results by Ma et al. showed that the double peak
structure of final-state dihadron correlation can also be
generated when a back jet traverses the dense core due to the
shadowing effect of the core~\cite{ma_jet}. In this picture, the
dense core shadowing effect is induced by the transverse expansion
or  radial flow of colliding system. When partons traverse through
a dense core region, they will suffer stronge interactions with
the medium, which will lead to extensive diffusions and particle
redistributions. In the event-averaged case when jet traverses the
dense medium, the double peak structure could be due to the sum of
deflected and/or cone-like jets. This dense core shadowing effect
is also related to path length dependence of jet-medium
interactions. The longer path length direction coincides with the
direction into the dense core, jet particles are blocked strongly
due to the shadowing effect. The asymmetry in the path length
distribution for particles interacting with the dense core medium
will result in the asymmetry in the dihadron correlation.

\begin{figure}[htbp]
\resizebox{8.6cm}{!} {\includegraphics{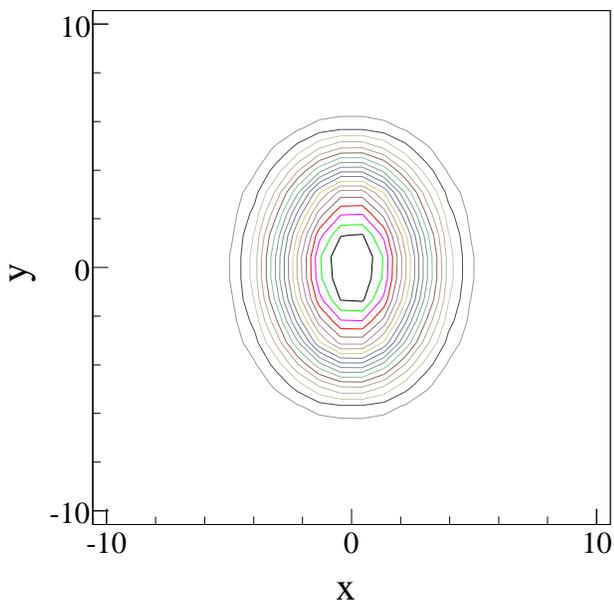}}
\vspace{-0.8cm} \caption{(Color online) The initial parton
distribution in transverse plane from AMPT model simulation of
Au+Au collisions at $\sqrt{s_{NN}}$ = 200 GeV
for 40$\%$ centrality. }  \label{initial_space_figure} 
\end{figure}

\begin{figure}[htbp]
\resizebox{8.6cm}{!} {\includegraphics{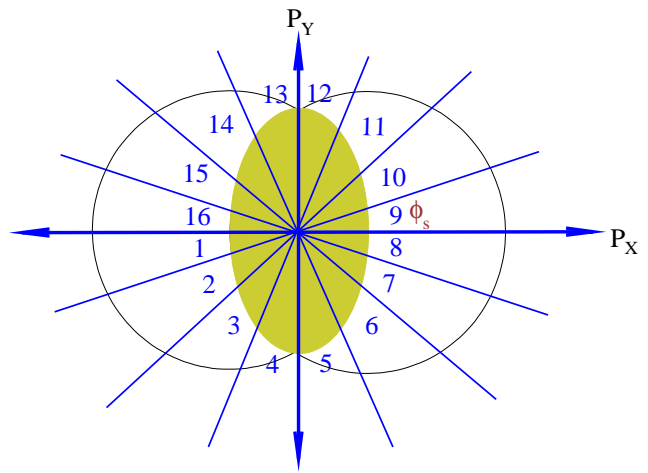}}
\vspace{-0.8cm} \caption{(Color online) Schematic illustration of
16 slices of trigger particle azimuthal angle relative to the
event plane, $\phi_{s}$ = $\phi_{T}$ - $\Psi_{EP}$. }
 \label{16show_figure} 
\end{figure}

\begin{figure}[htbp]
\resizebox{8.6cm}{!}{\includegraphics{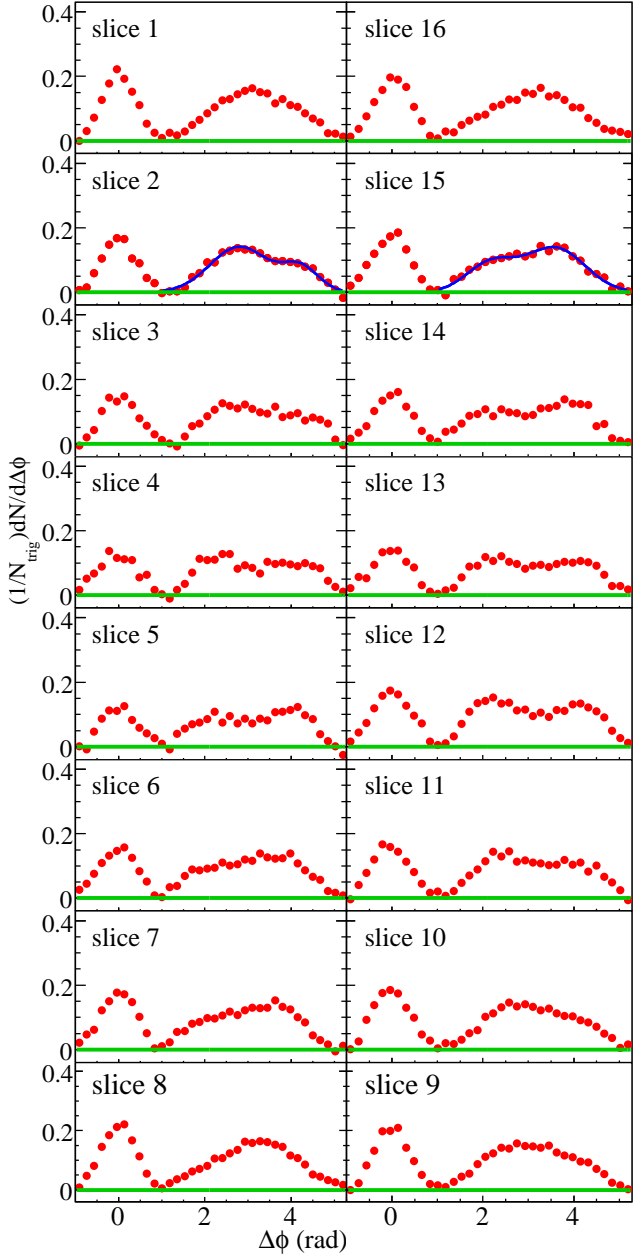}}
\vspace{-0.8cm} \caption{(Color online) Background subtracted
dihadron correlations with trigger particles in 16 slices of
azimuthal angle relative to the event plane,
 $\phi_{s}$ = $\phi_{T}$ - $\Psi_{EP}$. The data are from the melting AMPT model simulations with hadron rescattering on Au+Au collisions at $\sqrt{s_{NN}}$ = 200 GeV
for 20-60$\%$ centrality.
($p_{T}$ window cut in model: $2.5 <p_{T}^{trig}< 6$ GeV/$c$ and
 $0.15 <p_{T}^{asso}< 3$ GeV/$c$). Both the trigger and associated particles
are restricted within $|\eta|<1$. The solid line in Slice 2 and 15
represents the two-Gaussians fit for the away side dihadron
correlation.} \label{16_figure} \vspace{-0.5cm}
\end{figure}

To further investigate the path-length effect, we compared
correlation functions from mid-central collisions with central
collision results. We show dihadron azimuthal angle correlation
functions for 0-10$\%$ centrality and 20-60$\%$ centrality from
AMPT model simulations in Figure 4. We divide the first quadrant
of initial phase space region  into 6 equal slices based on
$\phi_{s}$ and in order to increase the statistics, we fold the
other three quadrants into the first quadrant. Significant
differences in the correlation functions between these two
centrality bins become apparent. For mid-central collisions, we
observe an evolution from a single peak to double peak structure.
But for central collisions, the correlation functions show little
variation with $\phi_{s}$ and double peak structure is present for
all slices. This can be explained that for central collisions, the
overlap region of the two colliding nuclei is almost spherical and
the dimensions of the dense medium are such that both in-plane and
out-of-plane path length are long enough to generate a splitting
structure. We conclude that in order to generate a double peak
structure, a dense medium of sufficiently large size is needed
\cite{Song09}. Our simulation results indicate that the onset of
the splitting structure corresponds to $\phi_{s}\sim45^{\circ}$
for 20-60\% centrality. The associated particle yields per
triggered particle for central collisions is larger than that for
mid-central collisions on both near and away side.

\begin{figure}[htbp]
\resizebox{8.6cm}{!}
{\includegraphics{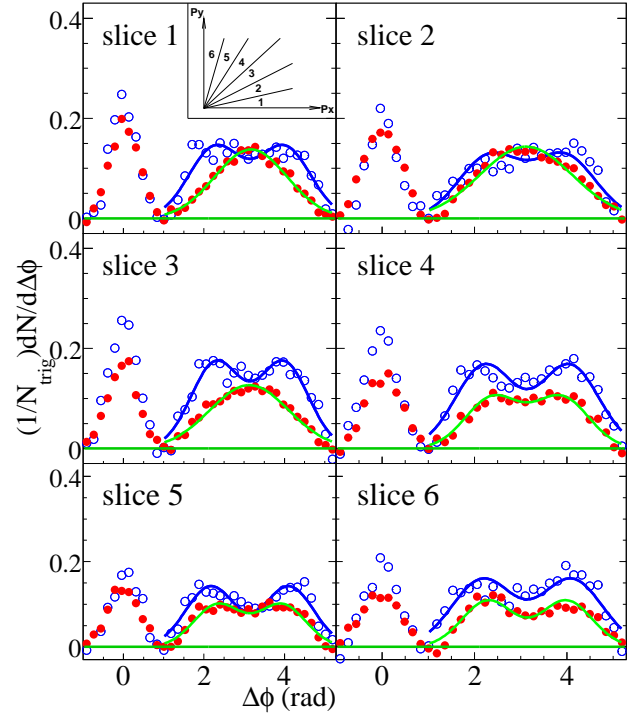}}\vspace{-0.5cm}
\caption{(Color online) Background subtracted dihadron
correlations for trigger particles of $2.5 <p_{T}^{trig}< 6$
GeV/$c$ and associated particles of $0.15<p_{T}^{asso}< 3$ GeV/$c$
from AMPT model simulation of Au+Au collisions at $\sqrt{s_{NN}}$
= 200 GeV with trigger particles in 6 equal slices of azimuthal
angle from the event plane in the first quadrant for 0-10$\%$
collision centrality (open circles) and 20-60$\%$ collision
centrality (solid circles). The $\eta$ window is restricted within
$|\eta|<1$. Lines represent the two-Gaussians or one-Gaussian fit
for away-side correlation.} \label{comparison_figure}
\vspace{-0.2cm}
\end{figure}

In what follows, we will extract a few parameters from the
dihadron angular correlation functions to describe the away-side
correlation shape and amplitude. Systematic comparisons of these
parameters between experimental results and theoretical
calculations will provide insight on the response of the medium to
the jet energy loss, and may provide new constraints for
understanding competing mechanisms for energy transport.

Figure 5 presents the AMPT model calculations of the trigger
direction dependence of the away side broadening from Au+Au
collisions at $\sqrt{s_{NN}}$ = 200 GeV. For central
collisions, the away-side width $W_{rms}$ grows very slightly with
$\phi_{s}$. For 20-60$\%$ central collisions, however, the away-side
$W_{rms}$ grows distinctly with trigger direction $\phi_{s}$ which
qualitatively agrees well with the experimental measurement~\cite{aoqi}. For
$\phi_{s}$ close to $0$ $ rad$, the $W_{rms}$ for 0-10$\%$ Au+Au
collisions is already large enough which represents a remarkable
broadening of the away-side correlation function, while that for the 20-60$\%$ Au+Au
collisions is relatively small corresponding to a narrow correlation.
This is consistent with expected dependence on the traversing path length
in the reaction plane direction
between these two centrality bins.
For $\phi_{s}$ close to $\pi/2$
$ rad$, the $W_{rms}$ of these two centrality bins are very much
close to each other. This can be attributed to the fact that the path length
perpendicular to the reaction plane is not much different between
these two centrality bins. The path length variation with the
trigger particle orientation in the central collisions is
relatively small compared to that in the mid-central collisions.
Accordingly, the away side $W_{rms}$ for central collisions varies
little with $\phi_{s}$ and is larger than that for mid-central
collisions at all trigger particle orientations.

\begin{figure}[htbp]
\resizebox{8.6cm}{!}{\includegraphics{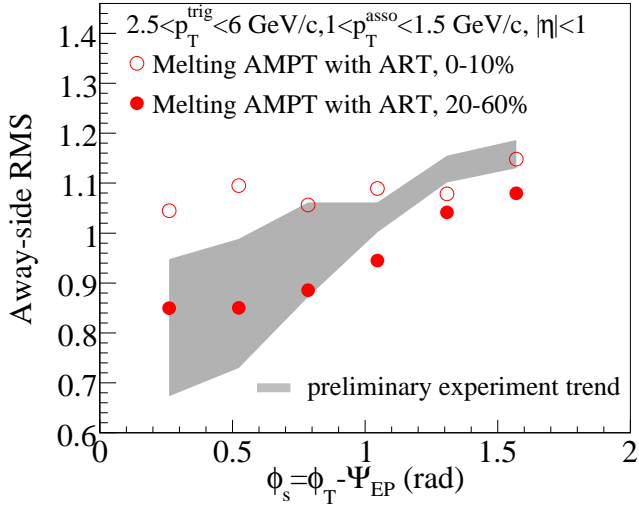}}
\vspace{-0.8cm} \caption{(Color online) The away-side $W_{rms}$ of
the dihadron correlation function versus the trigger particle
azimuthal angle relative to the event plane, $\phi_{s}$, from the
AMPT model calculation with $p_{T}$ window cut ($2.5
<p_{T}^{trig}< 6$ GeV/$c$ and $1 <p_{T}^{assoc}< 1.5$ GeV/$c$ ) in
0-10$\%$ (open circles) and 20-60$\%$ (solid circles) Au+Au
collisions at $\sqrt{s_{NN}}$ =200 GeV. The $\eta$ window is
restricted within $|\eta|<1$. The preliminary experimental trend
is for 20-60\% centrality~\cite{aoqi}. } \label{$R_{rms}$_figure}
\vspace{-0.2cm}
\end{figure}

Figure 6 shows the trigger particle azimuthal angle dependence of
the away-side splitting parameter $D$ (half distance between two
splitting peaks on away-side in dihadron azimuthal angle
correlation). For mid-central collisions, the splitting parameter
$D$ grows significantly with the trigger particle azimuthal angle
while for central collisions it grows only slightly. It can also
be attributed to different path-length variations between these
two centrality bins. We also show the trigger particle azimuthal
angle dependence of correlation yields for both near and away side
in Figure 7. On near-side, the yields decrease with trigger
azimuthal angle while on away-side, the yields increase.

\begin{figure}[htbp]
\resizebox{8.6cm}{!}{\includegraphics{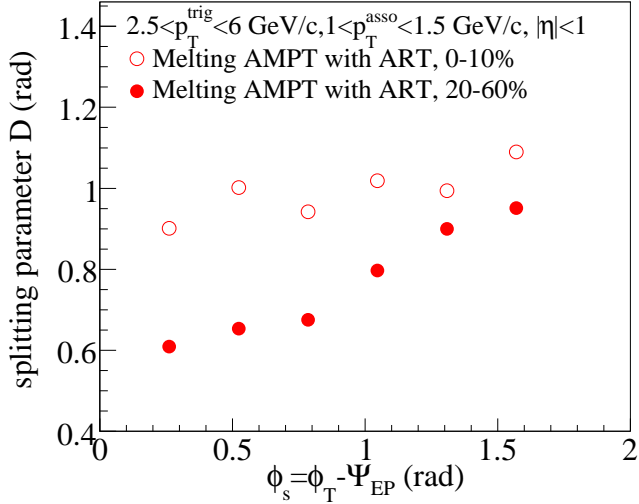}}
\vspace{-0.8cm}
 \caption{(Color online) The splitting parameter $D$ as a function of trigger
particle azimuthal angle from the event plane for trigger hadrons
of $2.5 <p_{T}^{trig}< 6$ GeV/$c$ and associated hadrons of $1
<p_{T}^{assoc}< 1.5$ GeV/$c$ from AMPT model simulations for Au+Au
0-10$\%$ (open circles) and 20-60$\%$ (solid circles) collisions
at $\sqrt{s_{NN}}$ = 200 GeV. } \label{D_figure} \vspace{-0.5cm}
\end{figure}

Figure 8 presents the $\Delta\phi$ dependence of mean $p_{T}$ for
in-plane and out-of-plane associate particles. We have combined data
of the slices 1, 2, 7, 8, 9, 10, 15 and 16 into in-plane result and the
rest into out-of-plane result. On the near-side, the mean $p_{T}$ values for both
in-plane and out-of-plane agree well with each other and they
stay almost constant. On the away-side, they show different angular dependence.
The mean $p_{T}$ for in-plane stays almost flat while for
out-of-plane, the mean $p_{T}$ displays concave structure with the minimal value at
around $\pi$ rad. Previously published results indicate
 that for central collisions the concave
structure implies that harder associated hadrons prefer larger
angles with respect to the back-to-back direction and it may be
attributed to the average interaction length of the away-side
partons interacting with the medium is maximal at $\Delta\phi=\pi$ $rad$~\cite{transverse-momentum}.
Our results indicate that
for mid-central collisions, the path-length on away-side for
in-plane is relatively small, and it varies very little with
$\Delta\phi$. For out-of-plane, however, partons on the away-side direction
will travel through a long enough path length in the medium leading to
a significant energy loss;
partons emitted away from the back-to-back direction will
traverse through relatively smaller path length with a correspondingly
smaller energy loss. The path-length effect plays an
important role on the structure of mean $p_{T}$ for the away side
particles.

\begin{figure}[htbp]
\resizebox{8.6cm}{!}{\includegraphics{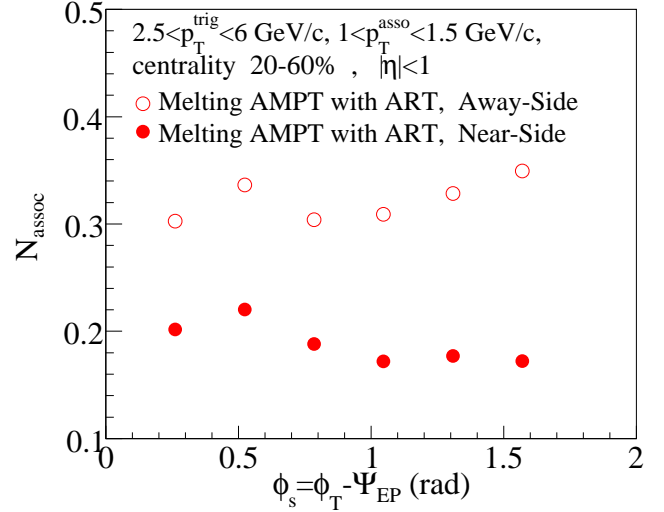}}
\vspace{-0.8cm}
 \caption{(Color online) The yields of the dihadron correlation
function on near (solid circles) and away side (open circles)
versus the trigger particle azimuthal angle from the event plane,
$\phi_{s}$, with $p_{T}$ window cut ($2.5 <p_{T}^{trig}< 6$
GeV/$c$ and $1<p_{T}^{assoc}< 1.5$ GeV/$c$) in 20-60$\%$ Au+Au
collisions at $\sqrt{s_{NN}}$ =200 GeV. The $\eta$ window is
restricted within $|\eta|<1$. } \label{Yields_figure}
\vspace{-0.5cm}
\end{figure}

\begin{figure}[htbp]
\resizebox{8.6cm}{!}{\includegraphics{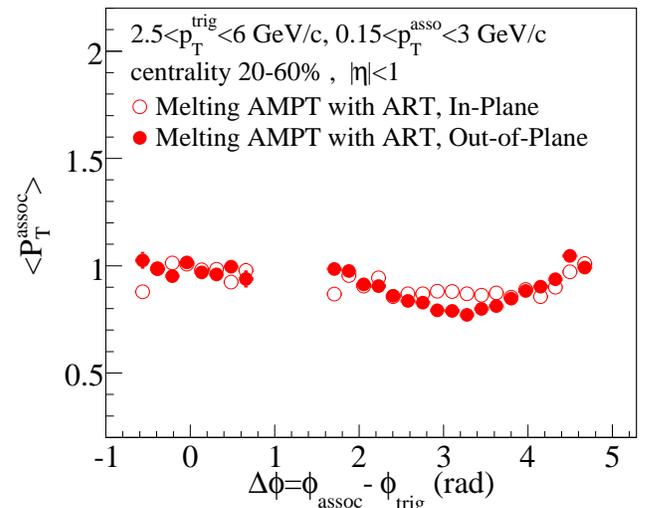}}
\vspace{-0.8cm}
 \caption{(Color online) The mean $p_{T}^{assoc}$ versus
$\Delta\phi$ ($\phi_{assoc}-\phi_{trig}$) for in- (open circles)
and out-of-plane (solid circles) in dihadron $\Delta\phi$
correlations with $p_{T}$ window cut ($2.5 <p_{T}^{trig}< 6$
GeV/$c$ and $0.15<p_{T}^{assoc}< 3$ GeV/$c$) in Au+Au collisions
(20-60$\%$) at $\sqrt{s_{NN}}$ =200 GeV. The $\eta$ window is
restricted within  $|\eta|<1$. }
\label{pt_in_out_figure} 
\end{figure}

\section{Conclusions}

In conclusion, dihadron azimuthal angle correlation functions with
respect to reaction plane have been studied in the framework of a
hybrid dynamical transport model, namely the AMPT model. In this
model, the energy loss is essentially from elastic parton-parton
scattering process, and the gluon radiation energy loss mechanism
has not been included yet. In order to match the experimental
observations, a large elastic cross section has to be used. With
the AMPT model simulations we observed the emerging of the
splitting structure in the away-side dihadron correlations from
Au+Au collisions at 200GeV/c. We found that the structure in the
azimuthal angle correlation functions has a strong dependence on
the trigger particle orientation relative to the reaction plane,
which is consistent with the observation from the  STAR
experiment. The correlation amplitude on the near-side slightly
decrease from the in-plane to the out-of-plane direction, while
the shapes of the away-side angular correlation functions evolve
from a single peak to double peaks. For the triggered particles
emitting from middle azimuthal angles relative to the reaction
plane, the corresponding away-side correlation function is
asymmetric which indicates that the shape of the correlation shall
be sensitive to the traversing path length. By comparing results
in the top 0-10$\%$ and 20-60\% collision centrality bins, we
found that the away-side path length in the reaction plane
direction in 20-60$\%$ Au+Au collision was not long enough to
generate a double peak structure while that in the top 0-10$\%$
collision it was long enough to do that. While in the out-of-plane
direction, the correlation structure seems to be not very
different since the path length is similar between these two
centrality bins along this direction. We extracted a few
parameters from correlation functions and discussed the possible
underlying mechanisms for the splitting structure. Our results
indicate that the away-side $W_{rms}$ and splitting parameter $D$
increase with $\phi_{s}$ for 20-60$\%$ centrality while for
0-10$\%$ centrality they change very slightly. This observation
underscores the importance of the path-length effect when the
away-side associated particles traverse the medium. The yields on
near-side decrease slightly with $\phi_{s}$ while on the away-side
they increase. We also investigate the $\Delta\phi$ dependence of
associated particles mean $p_{T}$. The dip structure of mean
$p_{T}$ as a function of $\Delta\phi$ along the out-of-plane
direction is more prominent than that along the in-plane
direction. These results could shed light on path-length effect of
elastic collision energy loss for high transverse momentum
particles propagating through the hot and dense medium. Strong
parton cascade processes and resultant energy loss play important
roles in the interaction between the jet-like particles and medium
in the AMPT model. Furthermore, the preferential way to
reconstruct the reaction plane in experiment now is using the
$m$-th harmonic plane, of which the second harmonic plane, i.e.,
$v_2$-method, is often used ~\cite{aoqi}. If one uses the first
harmonic plane ($v_1$ method), it will be possible to distinguish
between structures of different trigger particle emission regions,
thus allowing to separate different path length effect in
experiment. We note that the present findings are mostly relevant
for the elastic energy loss and we do not know contributions from
gluon radiation loss and their consequence on dihadron
correlations yet in a dynamical model simulation. Nevertheless,
our studies indicate that path length effect in jet medium
interactions is very important and should be thoroughly
investigated both experimentally and theoretically.

\section*{Acknowledgements}
This work was supported in part by the National Natural Science
Foundation of China under Grant No. 10610285, 10775167, 29010702,
10705044, the Knowledge Innovation Project of the Chinese Academy
of Sciences under Grant No. KJCX2-YW-A14, the Shanghai Development
Foundation for Science and Technology under Grant Nos.
09JC1416800. And we thank Information Center of Shanghai Institute
of Applied Physics of Chinese Academy of Sciences for using
PC-farm.


\end{document}